\documentclass[iop]{emulateapj}
\usepackage{amsmath, amssymb}
\usepackage{bm, graphicx}
\usepackage[counterclockwise]{rotating}
\usepackage{hyperref}
\usepackage{multirow}
\hypersetup{
	colorlinks = true,
	linkcolor=blue,
	citecolor=blue
}
\usepackage{etoolbox}

\makeatletter

\patchcmd{\NAT@citex}
{\@citea\NAT@hyper@{%
		\NAT@nmfmt{\NAT@nm}%
		\hyper@natlinkbreak{\NAT@aysep\NAT@spacechar}{\@citeb\@extra@b@citeb}%
		\NAT@date}}
{\@citea\NAT@nmfmt{\NAT@nm}%
	\NAT@aysep\NAT@spacechar\NAT@hyper@{\NAT@date}}{}{}

\patchcmd{\NAT@citex}
{\@citea\NAT@hyper@{%
		\NAT@nmfmt{\NAT@nm}%
		\hyper@natlinkbreak{\NAT@spacechar\NAT@@open\if#1\else#1\NAT@spacechar\fi}%
		{\@citeb\@extra@b@citeb}%
		\NAT@date}}
{\@citea\NAT@nmfmt{\NAT@nm}%
	\NAT@spacechar\NAT@@open\if#1\else#1\NAT@spacechar\fi\NAT@hyper@{\NAT@date}}
{}{}

\makeatother

\newcommand{\MYhref}[3][blue]{\href{#2}{\color{#1}{#3}}}

\slugcomment{The Astrophysical Journal, \MakeLowercase{000:00 (13pp), \today}}

\shorttitle{Role of Downflows in Solar Near-Surface Shear }
\shortauthors{Matilsky et al.}

\begin{document}


\title{The Role of Downflows in Establishing Solar Near-Surface Shear}


\author{Loren I. Matilsky\altaffilmark{1}, Bradley W. Hindman and Juri Toomre}
\affil{JILA \& Department of Astrophysical and Planetary Sciences, University of Colorado, Boulder, CO 80309-0526, USA}

\altaffiltext{1}{loren.matilsky@colorado.edu}


\begin{abstract}
The dynamical origins of the Sun's tachocline and near-surface shear layer (NSSL) are still not well understood. We have attempted to self-consistently reproduce a NSSL in numerical simulations of a solar-like convection zone by increasing the density contrast across rotating, 3D spherical shells. We explore the hypothesis that high density contrast leads to near-surface shear by creating a rotationally unconstrained layer of fast flows near the outer surface. Although our high-contrast models do have near-surface shear, it is confined primarily to low latitudes (between $\pm15^\circ$). Two distinct types of flow structures maintain the shear dynamically: rotationally \textit{constrained} Busse columns aligned with the rotation axis and fast, rotationally \textit{unconstrained} downflow plumes that deplete angular momentum from the outer fluid layers. The plumes form at all latitudes, and in fact are more efficient at transporting angular momentum inward at high latitudes. The presence of Busse columns at low latitudes thus appears essential to creating near-surface shear in our models. We conclude that a solar-like NSSL is unobtainable from a rotationally unconstrained outer fluid layer alone. In numerical models, the shear is eliminated through the advection of angular momentum by the meridional circulation. Therefore, a detailed understanding how the solar meridional circulation is dynamically achieved will be necessary to elucidate the origin of the Sun's NSSL. 
\end{abstract}



\keywords{convection --- turbulence  --- Sun: interior --- Sun: rotation --- Sun: kinematics and dynamics}


\section{Introduction}
\begin{figure}
	\hspace*{1cm}
	\includegraphics[width=0.35\textwidth, height=0.3\textwidth]{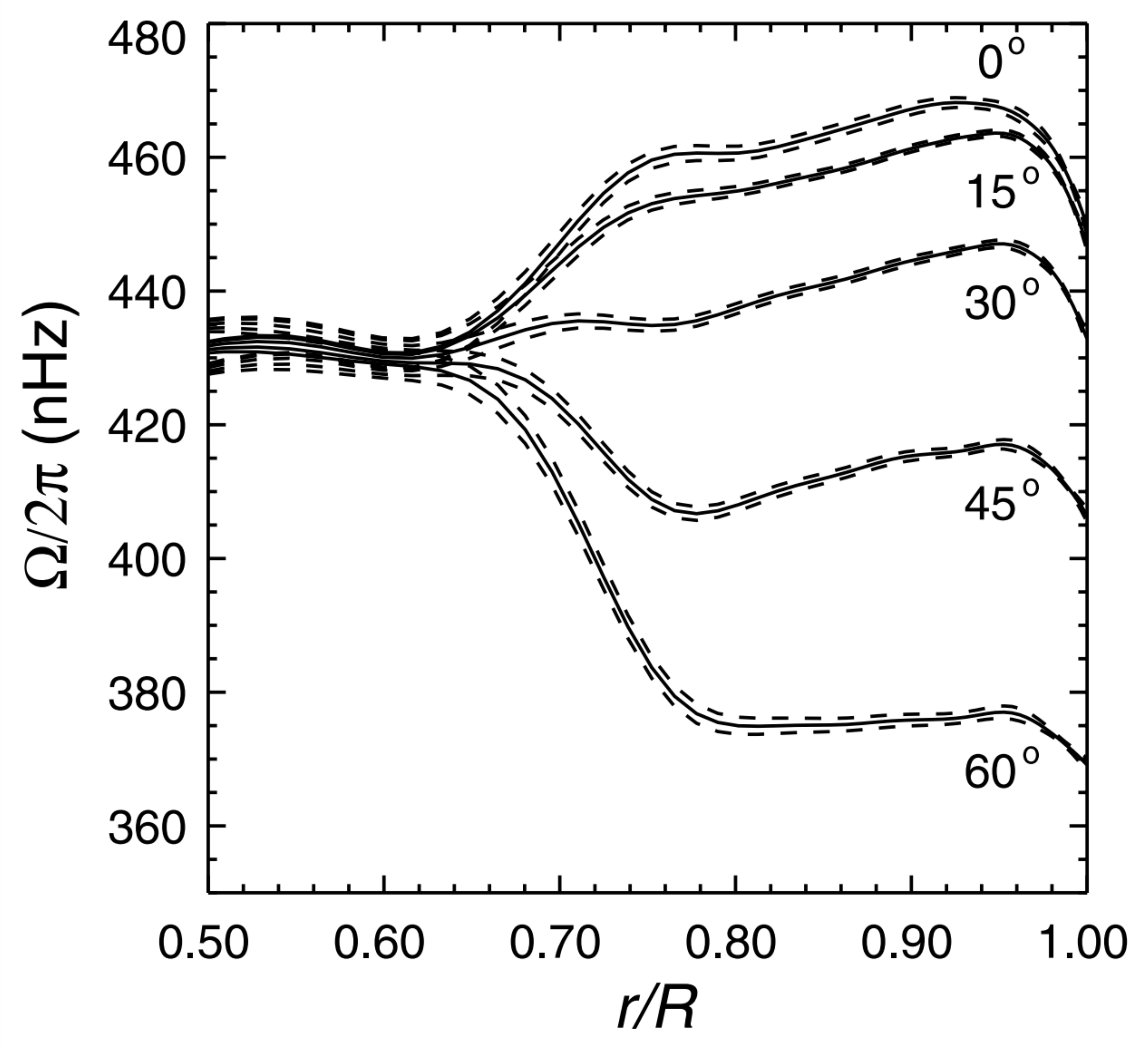}
	\caption{Time-averaged cyclic rotation rates $\Omega/2\pi$ obtained from inversion of GONG frequency splittings, plotted as a function of the fractional solar radius at different latitudes. Dashed lines represent 1$\sigma$ error bounds for a single inversion. From Howe, R., Christensen-Dalsgaard, J., Hill, F., et al., 2000, \href{http://science.sciencemag.org/content/287/5462/2456.long}{Science}, \MYhref[blue]{http://adsabs.harvard.edu/abs/2000Sci...287.2456H}{287, 2456}. Reprinted with permission from AAAS. Data from Global Oscillation Network Group/National Solar Observatory/AURA/NSF. Courtesy Dr. Rachel Howe. \label{fig:gongcut}}
\end{figure}

Helioseismology has revealed the presence of two boundary layers of shear at the top and bottom of the solar convection zone (CZ). In the \textit{tachocline} at the bottom, strong differential rotation in the CZ transitions sharply to solid-body rotation in the radiative zone below (see Figure \ref{fig:gongcut}). At the top of the CZ, there is a 5\% reduction in rotation rate with increasing radius over a depth of $\sim$35 Mm, which is largely a uniform feature at all latitudes. This latter feature is known as the \textit{near-surface shear layer} (NSSL). Both boundary layers may play a significant role in the solar dynamo, since rotational shear creates toroidal magnetic field from poloidal field through the $\Omega$-effect. However, the dynamical origins of these boundary layers are still not well understood.

 The strong solar differential rotation (from Figure \ref{fig:gongcut}, the equator rotates  about 30\% faster than the poles) is believed to be the result of convectively driven Rossby waves, which manifest in the solar convection zone as \textit{Busse columns}, also known as \textit{banana cells} in the literature (e.g., \citealt{Gilman72}; \citealt{Busse02}; \citealt{Brun02}; \citealt{Nelson18}). Busse columns are convective rolls of fluid aligned with the rotation axis. Each roll has a cross-sectional tilt in the equatorial plane such that upflows move prograde and downflows move retrograde, the net result being the transport of angular momentum away from the rotation axis (see Figure 6 of \citealt{Busse02}). The Busse columns thus tend to spin up the equator (which is far from the rotation axis) compared to the poles.
 
 By contrast, in the NSSL (see Figure \ref{fig:gongcut}), the surface layers rotate 5\% \textit{slower} than the layers just below.  \citet{Foukal75} hypothesize the following explanatory mechanism for the formation of the NSSL: fluid particles conserve their specific angular momentum $\mathcal{L}$ in the outermost layers of the Sun as they move in the radial direction. Thus, for a steady-state system, the angular momentum profile $\mathcal{L}(r)$ should be constant with radius. Since specific  angular momentum is related to the local fluid rotation rate $\Omega$ through $\mathcal{L} = \Omega r^2\sin^2\theta$, this would imply $\Omega \propto1/r^2$ along a radial line in the outermost fluid layers---a decrease in rotation rate with radius.

In order to successfully homogenize angular momentum, the flows need to be free from rotational constraint so that they do not get captured by Busse column rolls. The degree of rotational constraint is parameterized by the convective Rossby number $\rm{}Ro_c$, which is the ratio of rotational period to convective overturning time. Thus, according to \citet{Foukal75}, the Sun must possess a region near the outer surface in which 
\begin{equation}\label{eq:highRoc}
\rm{Ro_c} \gtrsim 1.
\end{equation}
The Rossby number may also be written $\rm{Ro_c}$ $= v^\prime/2\Omega_0L$, where $v^\prime$ is a typical velocity of the flow structure, $L$ is its typical length-scale and $\Omega_0$ is the frame rotation rate. Thus, a \textit{rotationally unconstrained} fluid structure is one that is both fast and small-scale. In the models explored here, we find that only the downflows (in particular, structures we call \textit{downflow plumes}) are sufficiently fast and small-scale to be rotationally unconstrained.

In this work, we attempt to reproduce a solar-like NSSL using 3D spherical-shell convection models that self-consistently generate both types of flows (Busse columns and downflow plumes), using Newtonian diffusivities and simple boundary conditions. We achieve a plume-dominated layer near the outer surface by increasing the density contrast across the spherical shell while keeping the density at the inner surface fixed. This decreases the density scale-height (which is a good representative length-scale of the convection) in the outer layers of the shell, and also accelerates downflows to higher speeds through the buoyancy force. Both effects serve to increase the Rossby number near the outer surface. We will find, however, that although this Rossby-number transition is indeed achieved, it is insufficient to create a solar-like NSSL. In our models, this is because of the detailed structure of the meridional circulation profile, thus making the NSSL an inherently \textit{global} problem. 

In Section \ref{sec:num}, we present the mathematical equations that are solved in our models and describe the parameter space we explore. In section \ref{sec:global_structure}, we discuss the global character of the flows achieved, both instantaneously and averaged over time. Sections \ref{sec:BusseColumns} and \ref{sec:DownflowPlumes} deal with the structure and evolution of Busse columns and downflow plumes, respectively.  In Section \ref{sec:torque}, we discuss the dynamical balance of torques in our models and its relation to the simulated features of near-surface-shear. In section \ref{sec:amom_flux}, we examine in detail the Reynolds stress from Busse columns and downflow plumes, which manifest in the separate upflow- and downflow-contributions to the angular momentum flux.  In Section \ref{sec:disc}, we discuss our results in the general context of meridional force balance.

\begin{sidewaysfigure*}
	\vspace*{10cm}
	\hspace*{-1cm}
	\includegraphics[width=1.05\textwidth]{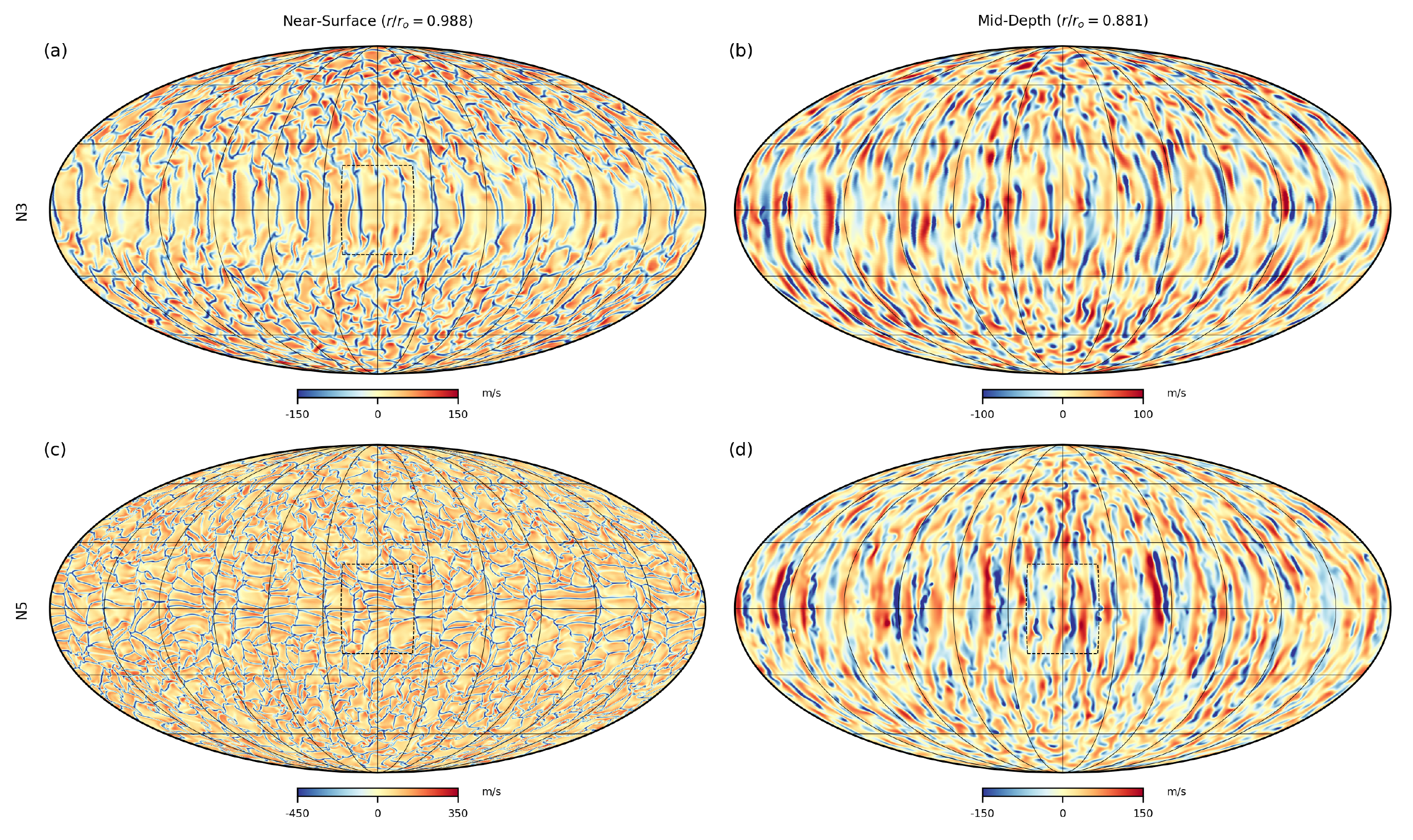}
	\caption{(\textit{a})---(\textit{d}): Mollweide projection of radial velocity $v_r$ on spherical surfaces for cases N3 and N5. Red tones indicate upflows ($v_r>0$), while blue flows indicate downflows ($v_r<0$). The spherical surfaces are at two radii, one at a near-surface layer 5\% of the way from top to bottom ($r/r_o=0.988$) and one at a layer mid-way through the domain ($r/r_o=0.881$). In panel (\textit{c}), the colorbar is binormalized to highlight the asymmetry in the amplitudes of upflows vs. downflows near the outer surface of case N5. The dashed $40^\circ\times40^\circ$ box at the center of each of panels (\textit{a}), (\textit{c}) and (\textit{d}) show the regions that are magnified in later figures.}
	\label{fig:sslice}
\end{sidewaysfigure*}

\section{Numerical Model}\label{sec:num}
We numerically evolve a rotating, stratified shell of fluid representative of the convection zone of a star. We use the MHD code \textit{Rayleigh} 0.9.1  (\citealt{Featherstone16}---hereafter, FH2016; \citealt{Matsui16}, \citealt{Featherstone18}), published under the GPL3 license. The computational domain of the layer consists of a spherical shell with inner radius $r_i$ and outer radius $r_o$. We generally use the standard spherical coordinates ($r,\theta,\phi$) and corresponding unit vectors ($\hat{\bm{e}}_r$, $\hat{\bm{e}}_\theta$, $\hat{\bm{e}}_\phi$), although for some topics we also use the cylindrical coordinates $(\lambda, \phi, z) = (r\sin\theta, \phi, r\cos\theta)$ and unit vectors ($\hat{\bm{e}}_\lambda$, $\hat{\bm{e}}_\phi$, $\hat{\bm{e}}_z$). 

\textit{Rayleigh} makes use of the anelastic approximation (e.g., \citealt{Gough68}; \citealt{Gilman81}) to increase the maximum allowable time step by removing sound waves. The thermodynamic reference state is chosen to be temporally steady and spherically symmetric with adiabatic stratification (see \citealt{Jones11} for a complete description). We denote the pressure, density, temperature and entropy by $P$, $\rho$, $T$ and $S$, respectively. We use overbars on the thermodynamic variables to denote the fixed reference state and the lack of overbars to denote deviations from the reference state. The equations representing conservation of mass, momentum and energy are then given by (see FH2016)
\begin{align}
\nabla\cdot(\overline{\rho}\bm{v}) &=  0,
\label{eq:cont}
\end{align}
\begin{align}
\overline{\rho}\Bigg{[}\frac{\partial\bm{v}}{\partial t} + (\bm{v}\cdot\nabla)\bm{v}\Bigg{]} &= -2\overline{\rho}\bm{\Omega}_0\times\bm{v} \nonumber\\
  &-\overline{\rho}\nabla \Bigg{(}\frac{P}{\overline{\rho}}\Bigg{)} -\frac{\overline{\rho} S}{c_p}\bm{g}
+ \nabla\cdot \bm{D}
\label{eq:mom}
\end{align}
and
\begin{align}
\overline{\rho}\overline{T}\Bigg{[}\frac{\partial S}{\partial t} + \bm{v}\cdot\nabla S)\Bigg{]} &= \nabla\cdot\big{[}\kappa\overline{\rho}\overline{T}\nabla S)\big{]} \nonumber \\
 &+ 2\overline{\rho}\nu\Big{[}e_{ij}e_{ij} - \frac{1}{3}(\nabla\cdot\bm{v})^2\Big{]} + Q,
\label{eq:en}
\end{align}
respectively. Here $\bm{v} = (v_r, v_\theta, v_\phi)$ is the local fluid velocity (in spherical coordinates) in the rotating frame, $c_p$ is the specific heat at constant pressure, $\bm{g}$ is the local gravitational acceleration due to a solar mass $M_\odot$ at the origin and $\kappa$ is the thermometric conductivity. The expression for momentum conservation---Equation \eqref{eq:mom}---employs the Lantz-Braginsky-Roberts approximation, which is exact for adiabatic reference states \citep{Lantz92, Braginsky95}. Both the diffusivities $\nu$ and $\kappa$ are fixed constants in space for our models. $\bm{D}$, $e_{ij}$ and $\delta_{ij}$ refer to the standard Newtonian viscous stress tensor, rate-of-strain tensor and Kronecker delta, respectively. In Equation \eqref{eq:en}, the standard summation convention for repeated indices is used. The internal heating function, given by the negative divergence of the radiative energy flux ($Q=-\nabla\cdot\bm{F}_{\rm{rad}}$), is chosen to have a fixed radial profile $Q(r)=\alpha[\overline{P}(r) - \overline{P}(r_o)]$, with the normalization constant $\alpha$ chosen such that a solar luminosity $L_\odot$ is forced through the domain (see FH2016). This prescription for the internal heating function coincides well with the radiative heating calculated by more sophisticated solar models (see model $S$ described in \citealt{Dalsgaard96}, for example). 

The equation set is closed with an equation of state for the thermodynamic variables, which consists of a perfect gas subject to small thermodynamic perturbations about the adiabatic reference state: $\rho/\overline{\rho} = P/\overline{P} - T/\overline{T} = P/\gamma\overline{P} - S/c_p$. Here, $\gamma$ is the ratio of specific heats.

We adopt boundary conditions on the velocity field to conserve angular momentum and mass. These are the stress-free and impenetrability boundary conditions:
$v_r = (\partial/\partial r)(v_\theta/r)  = (\partial/\partial r)(v_\phi/r)  = 0$ at $r=r_i,\ r_o$. The conditions on the entropy are such that heat does not enter through the inner boundary and there is constant entropy on the outer boundary: $\partial S/\partial r = 0$ at $r=r_i$ and  $S = 0$ at $r=r_o$. The outer boundary condition on the entropy ensures that as the system equilibrates, a sharp gradient in the spherical mean of $S$---i.e., a thermal boundary layer---develops near the outer surface, such that a solar luminosity is ultimately carried out of the layer via thermal conduction. This stands in contrast to the solar convection zone, where the energy is ultimately carried out by radiative cooling.

In this work, we compare two models with different density contrasts. Each rotates at roughly 3 times the solar Carrington rate. The relevant model parameters are shown in Table \ref{tab:model}. Here, $N_\rho$ refers to the number of density scale heights across the domain. The density contrast from the inner to the outer surface (also shown in Table \ref{tab:model}) is related to the number of scale heights by $\overline{\rho}_i/\overline{\rho}_o = \exp(N_\rho)$, where $\overline{\rho}_i$ and $\overline{\rho}_o$ refer to the values of $\overline{\rho}$ at the inner and outer boundaries, respectively. The \textit{thermal diffusion time} $T_{\rm{diff}}$ estimates how long it takes for heat to diffuse across the full spherical shell. The \textit{averaging time} refers to the time interval used in the temporal averages of fluid quantities---e.g., differential rotation, meridional circulation and Reynolds stress. This interval coincides with the time from equilibration (total energy flux constant with radius) to the end of the simulation. All input parameters are fixed in the two models, except for $N_\rho$, which takes on the values 3 and 5. We refer to the resulting models as N3 and N5, respectively. 
\begin{table}
	\caption{Input Model Parameters for Cases N3 and N5.}\label{tab:model}
	\centering
	\begin{tabular}{r  l  l}
		\hline\hline
		Parameter & N3 & N5\\
		\hline
		$r_i$       & \multicolumn{2}{c} {5.000$\times10^{10}$ cm}\\
		$r_o$  & \multicolumn{2}{c}{6.586$\times10^{10}$ cm}\\
		$c_p$ & \multicolumn{2}{c}{3.500$\times10^8$ erg K$^{-1}$ g$^{-1}$}\\
		$\nu$ & \multicolumn{2}{c}{2.000$\times10^{12}$ cm$^2$/s}\\
		$\kappa$ & \multicolumn{2}{c}{2.000$\times10^{12}$ cm$^2$/s}\\
		$\gamma$  & \multicolumn{2}{c}{1.667} \\
		$\overline{\rho}_i$ & \multicolumn{2}{c}{0.1805 g/cm$^3$} \\
		$\Omega_0$ & \multicolumn{2}{c}{7.800$\times10^{-6}$ rad/s} \\
		$\Omega_0/2\pi$ & \multicolumn{2}{c}{1,241 nHz}\\
		$P_0\equiv 2\pi/\Omega_0$ & \multicolumn{2}{c}{9.323 days}\\
		$T_{\rm{diff}} \equiv (r_o-r_i)^2/\kappa$ & \multicolumn{2}{c}{1,456 days (3.985 yr)}\\
		\hline
		$N_\rho$ & 3.000 & 5.000\\
		$\overline{\rho}_i/\overline{\rho}_o$ & 20.10 & 148.4\\
		\hline
		\multirow{3}{*}{Averaging time} & 53.78 yr & 34.07 yr\\ 
		& 2,107 $P_0$  & 1,335 $P_0$ \\
		& 13.50 $T_{\rm{diff}}$ & 8.550 $T_{\rm{diff}}$ \\
		\hline
	\end{tabular}
	\tablecomments{All parameters are displayed to 4 significant digits.}
\end{table}

\section{Global Flow Properties}\label{sec:global_structure}
Figure \ref{fig:sslice} shows the radial velocity of the flow on spherical surfaces near the outer boundary and at mid-depth for cases N3 and N5. The pairs of alternating lanes of upflow and downflow parallel to the rotation axis indicate the locations of pairs of Busse columns. The columns are spaced according to an azimuthal degree of $m\sim30$ for both N3 and N5. This corresponds to Busse columns with azimuthal extent $r_o-r_i$, the thickness of the spherical shell.  At mid-depth, the columns have connectivity to their counterparts in the near-surface layers, but they are significantly slower and the associated upflow and downflow lanes are thicker. At high latitudes, there is less noticeable alignment of the fluid structures with the rotation axis, though this may in part be due to angular distortion effects of the spherical projection. We note that in the near-surface layers of case N5, there are downflow lanes oriented North-South (parallel to the rotation axis) \textit{and} lanes oriented East-West (orthogonal to the rotation axis). The places where the two types of lanes cross we call \textit{interstices}. We shall see that the interstices are sources of prominent \textit{downflow plumes}, which evolve independently from the Busse columns and transport angular momentum in the opposite direction. 

\begin{figure}
	\includegraphics{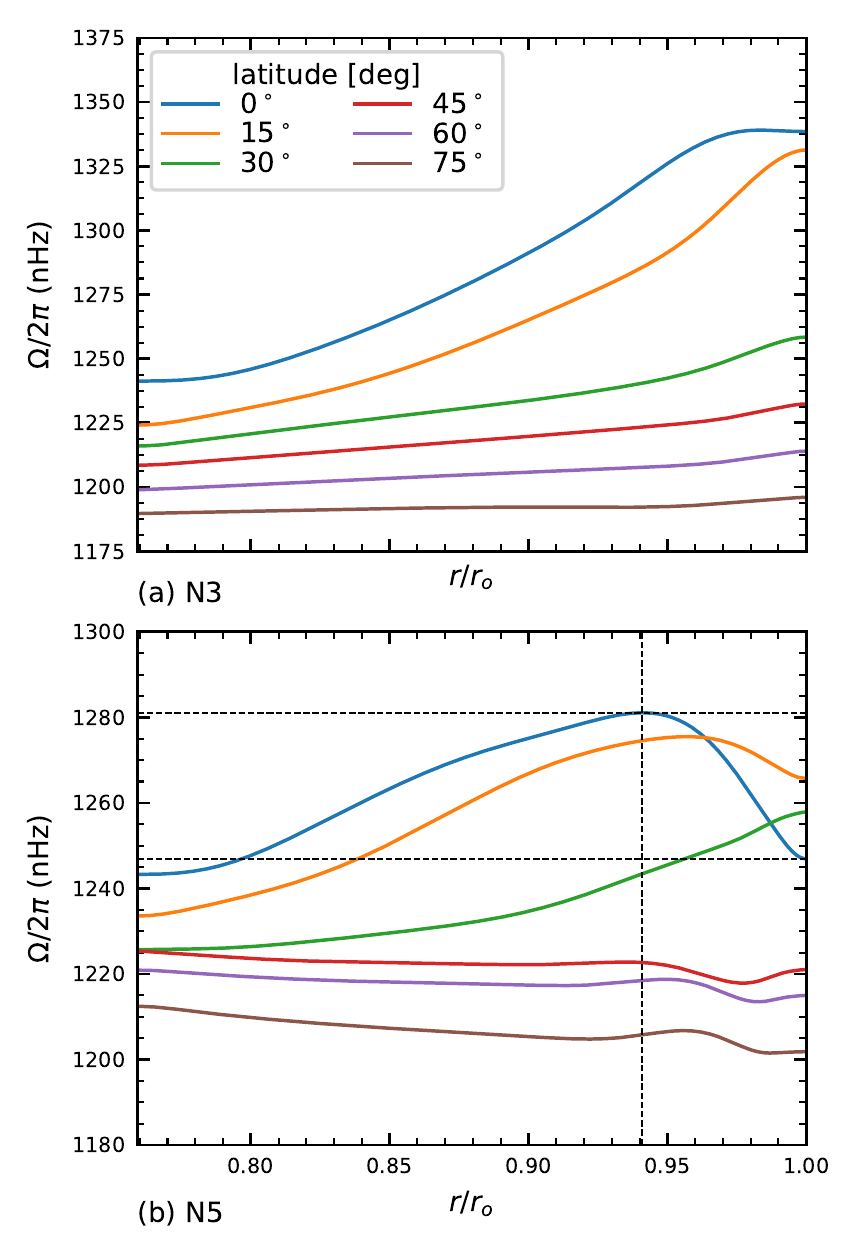}
	\caption{(\textit{a}): Temporally and azimuthally averaged rotation rate as a function of radius at various latitudes for case N3 (compare to the Sun's rotation in Figure \ref{fig:gongcut}). (\textit{b}): Rotation rate for case N5. The location of the prominent \textit{dip} in rotation rate in case N5 is indicated by the vertical dashed line at $r/r_o = 0.941$; its amplitude is indicated by the two horizontal dashed lines separated by 34.1 nHz. \label{fig:diffrot} }
\end{figure}

We turn next to the mean flow properties of our models. The average radial profile of rotation rate at various latitudes for cases N3 and N5 is shown in Figure \ref{fig:diffrot}. Case N3 exhibits the stronger differential rotation, with a variation of $\Delta\Omega=$ 155 nHz from equator to pole. This corresponds to a differential rotation fraction of $\Delta\Omega/\Omega_0 = 0.125$. As the  density contrast across the layer increases, the overall differential rotation $\Delta\Omega$ decreases, along with the differential rotation fraction. Case N5 has a differential rotation from equator to pole of only 81 nHz and a differential rotation fraction of $\Delta\Omega/\Omega_0=0.065$. We note that both these values of fractional differential rotation are significantly smaller than the value $\Delta\Omega_\odot/\Omega_\odot\approx0.3$ observed in the Sun. 

While the rotation rate in case N3 increases monotonically in both radius and latitude,  case N5 exhibits features reminiscent of near-surface shear. The most prominent of these is a \textit{dip} in the rotation rate at low latitudes near the outer surface: here, the rotation rate decreases with radius by about 2.7\%. The effect is comparable in magnitude to the Sun's NSSL, which is characterized by a roughly 5\% decrease in rotation rate near the top of the CZ. This low-latitude dip has been a robust feature of other work (e.g., \citealt{Brun02}; \citealt{Brandenburg07}; \citealt{Guerrero13}; \citealt{Gastine13}; \citealt{Hotta15}) and is also referred to as a rotation rate \textit{dimple} in the literature. At high latitudes in case N5, there are some signs of shear as well, although the overall effect is much weaker, corresponding to a reduction in angular velocity of only $\sim$$0.5\%$. Furthermore, the shear has both a negative and positive radial gradient. \citet{Hotta15} saw this phenomenon as well.

\begin{figure}
	\includegraphics{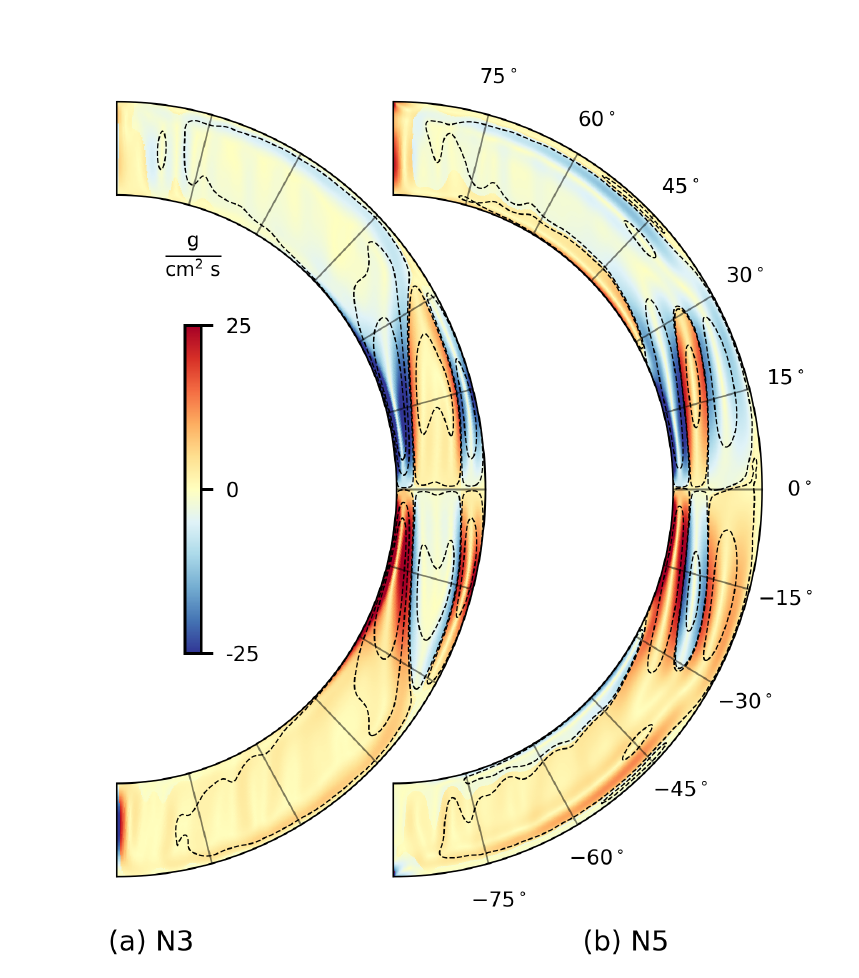}
	\caption{(\textit{a}) and (\textit{b}) show the vector-magnitude of the temporally and azimuthally averaged mass flux---$|\langle\overline{\rho}\bm{v}_m\rangle|$---in the meridional plane for cases N3 and N5. The vector-magnitude of the mass flux corresponds to the density of meridional circulation streamlines, several of which are denoted by the black, dashed contours. Here, $\bm{v}_m=v_r\hat{\bm{e}}_r + v_\theta\hat{\bm{e}}_\theta$ refers to the meridional part of the velocity. Red (positive values) indicates clockwise circulation, while blue (negative values) indicates anticlockwise circulation. \label{fig:massflux}}
\end{figure}
The magnitude of the zonally averaged mass flux for cases N3 and N5 is shown in Figure \ref{fig:massflux}. At low latitudes in the Northern hemisphere (for both cases), there are three cylindrically stacked circulation cells, the inner and outer cells being anticlockwise  and the sandwiched inner cell being clockwise. The central clockwise cell is large in case N3, but small in case N5, while the opposite is true for the anticlockwise cell near the outer surface. Furthermore, case N5 has two additional clockwise cells at high latitudes, one near the outer boundary and one near the inner boundary.  For both cases, the large anticlockwise cell at high latitudes is concentrated in a thin band near the outer surface, where there is strong poleward flow. In the Southern hemisphere for each case, the circulation patterns are the same, but with the clockwise-anticlockwise sense of each cell reversed.

\begin{figure}
	\includegraphics[width=0.5\textwidth]{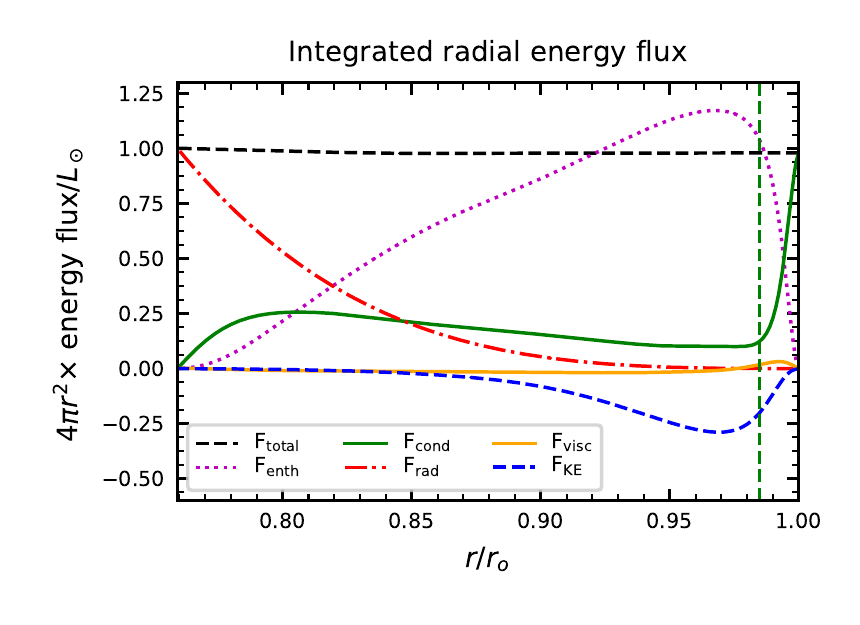}
	\caption{Decomposition of the spherically integrated radial energy flux for case N5, as described in FH2016. The location of the base of the thermal boundary layer is shown as the dashed vertical green line, chosen by eye to coincide with the near-surface local minimum (in radius) of the conductive energy flux. \label{fig:eflux_radial}}
\end{figure}

Figure \ref{fig:eflux_radial} shows the breakdown of radial energy fluxes (as defined in FH2016) for case N5. For the most part, four main fluxes contribute: the radiative flux $F_{\rm{rad}}$, the conductive flux $F_{\rm{cond}}$, the enthalpy flux $F_{\rm{enth}}$ (which represents the convective transport of heat) and the kinetic energy flux $F_{\rm{KE}}$. In the bottom layers ($r/r_o \lesssim 0.85$), energy is transported primarily by the radiative flux (with about 25\% of the energy being transported by the conductive flux). As the radiative flux decreases, the enthalpy flux begins to take over. Around $r/r_o\sim0.97$, the convective heat flux is dominant. Finally, near the outer surface, the boundary conditions on the velocity and entropy force all fluxes to vanish except for the conductive flux, which carries a solar luminosity out of the domain in a narrow thermal boundary layer. The extreme flatness of the total energy flux in case N5 indicates a mature time-averaged equilibrium.  
\begin{figure}
	\includegraphics{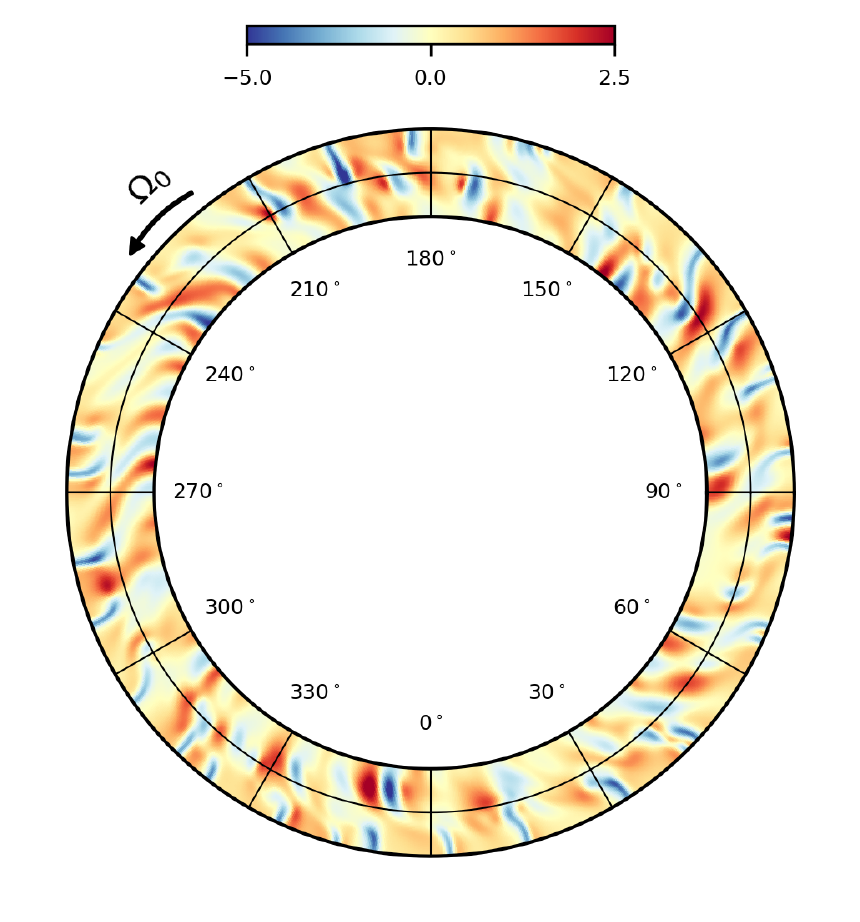}
	\caption{Instantaneous profile of the radial velocity $v_r$ on an equatorial cut of the computational domain for case N3. The view is from the North Pole, so the longitude $\phi$ increases in an anticlockwise sense. The radial velocity has been divided by its rms value at each radius and the colorbar is binormalized to show the asymmetry in the upflow and downflow speeds with respect to the rms. The radius midway through the layer at $r/r_o=0.881$ is marked by the central thin black circle. \label{fig:eqslice}}
\end{figure}

\section{Busse Columns}
\label{sec:BusseColumns}
We recall that \textit{Busse columns} are convective rolls of fluid aligned with the rotation axis. Adjacent rolls have opposite senses of spin, so that each columnar downflow lane traces the region in between two Busse-column rolls. In Figure \ref{fig:eqslice}, we show the equatorial cross section of radial velocity in the fluid layers for case N3. The prograde tilt of the downflow lanes is obvious: the portions of the lane close to the outer surface are at a higher longitude than the portions of the column close to the inner surface. As a general rule, the columns extend in depth all the way through the layer; however, several structures (especially the downflows in the upper half of the layer) only extend through half the layer or less. 

\begin{figure}
	\includegraphics{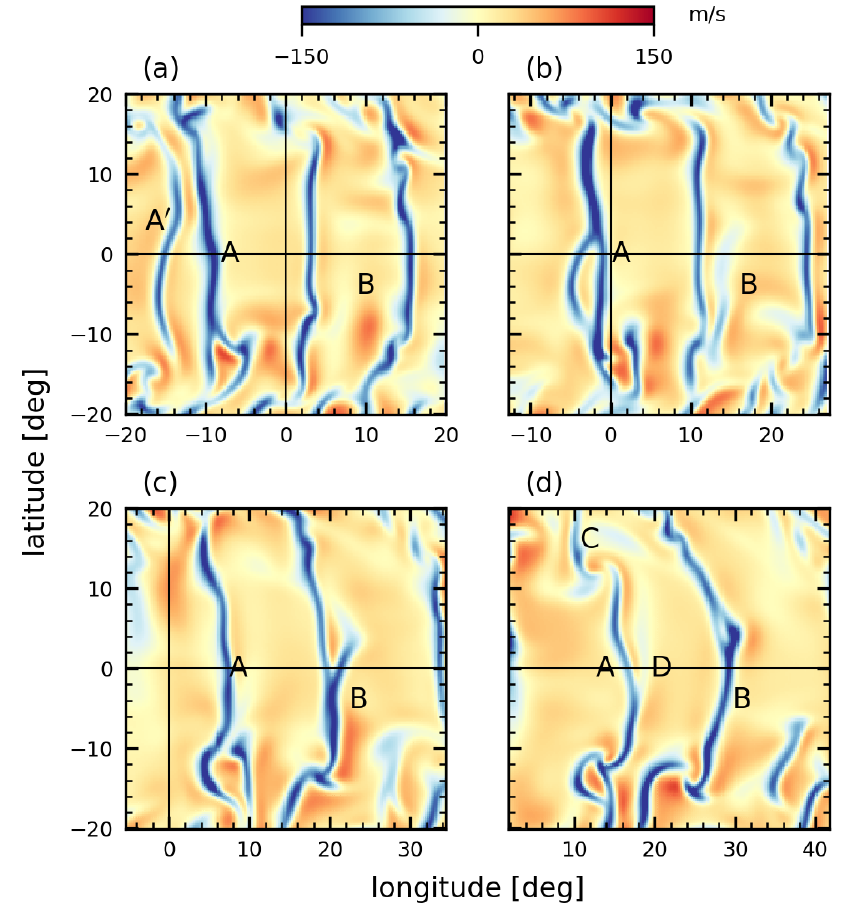}
	\caption{Temporal evolution of the near-surface radial velocity $v_r$ for case N3. Spherical surfaces are taken at 5\% depth ($r/r_o=0.988$). Each panel shows a 40$^\circ\times$  40$^\circ$ patch of the spherical surface centered at the equator, with successive patches equally spaced in time by a quarter of a rotation period. The frame of the patch is rotating at the local equatorial rotation rate in order to see the super-rotation of the columns. We label salient fluid structures discussed in the main text with capital letters. \label{fig:n3_3_patch_evolve}}
\end{figure}

Figure \ref{fig:n3_3_patch_evolve} shows the temporal evolution of the near-surface flow field in case N3. To better see the evolution, we magnify a 40$^\circ\times40^\circ$ patch centered at the equator, indicated by the dashed box in Figure \ref{fig:sslice}(\textit{a}). Some downflow lanes (which trace the regions in between adjacent Busse-column rolls) have been labeled with capital letters to indicate how they are advected and distorted by the flow. Although the downflow lanes are rather long-lived (lane A, for instance, maintains its structure for several rotation periods before getting absorbed by another lane), they are not simply advected passively by the flow. They frequently merge, disappear and reappear, indicating that no one Busse column lasts for a protracted interval. Furthermore, the lanes (and hence, the columns) only extend coherently to about $\pm15^\circ$; beyond this latitude range, the lanes move more slowly, eventually breaking off and joining the swirling small-scale flow at high latitudes. We note the differential rotation of Figure \ref{fig:diffrot}(\textit{a}) correspondingly occupies mainly the narrow latitude band between $\pm15^\circ$. Each lane is advected by several degrees over the whole rotation period, corresponding to a pattern speed of the Busse columns that is $\sim$40 m/s \textit{faster} than the background rotation rate. In other words, the Busse columns \textit{super-rotate} with respect to the background flow.

\section{Downflow Plumes}
\label{sec:DownflowPlumes}
\begin{figure}
	\includegraphics{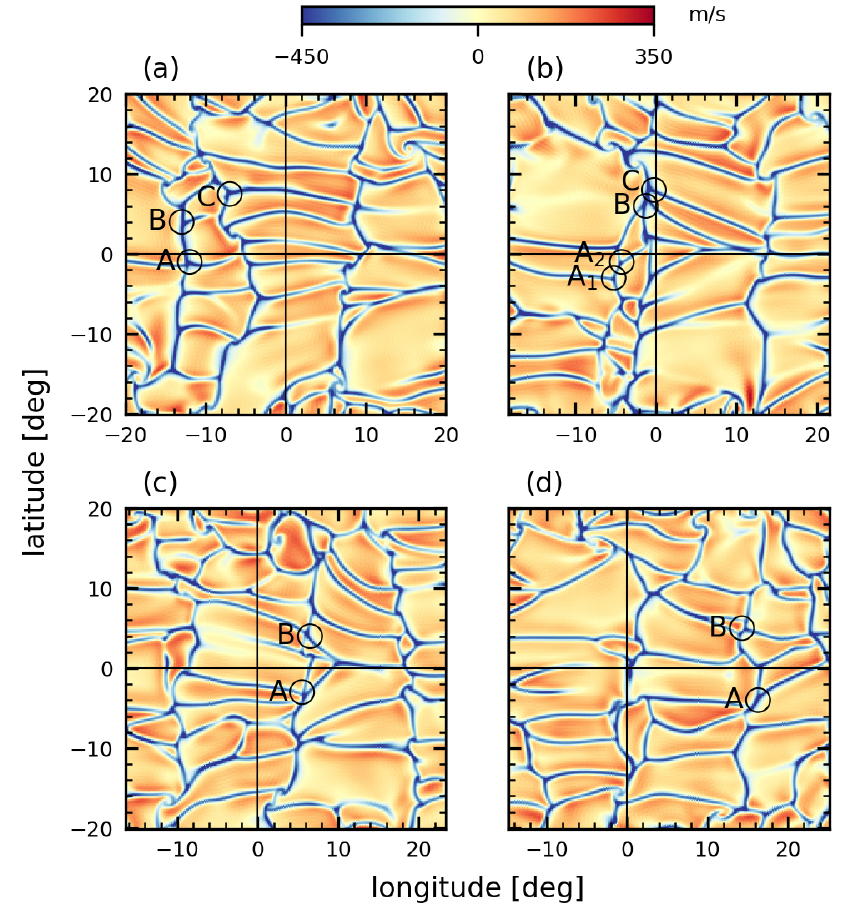}
	\caption{Similar to Figure \ref{fig:n3_3_patch_evolve}, but for case N5. Spherical surfaces are taken at $r/r_o=0.988$ and each panel shows a $40^\circ\times40^\circ$ patch of the spherical surface centered at the equator. The times of the panels correspond to those of Figure \ref{fig:n3_3_patch_evolve}. The frame of the patch is rotating at the local equatorial rotation rate. \label{fig:n5_3_patch_evolve}}
\end{figure}

 Figure \ref{fig:n5_3_patch_evolve} shows the evolution of the near-surface flow field for case N5, with several interstices (regions where the North-South and East-West downflow lanes cross) labeled by capital letters. We see that each upflow is surrounded by a more-or-less polygonal network of downflow lanes and thus may be regarded as a \textit{cell}. The cells are stacked in the axial direction such that there are 10 or so downflow lanes that connect throughout the whole domain North--South, analogous to the downflow lanes between pairs of Busse column rolls in case N3. On average, the interstices move in the prograde direction, indicating that they are super-rotating like the Busse columns, though more slowly. The interstices have much shorter lifetimes than the Busse columns, splitting up and merging several times over the course of a rotation period. In panel (\textit{b}), for example, interstice A has split into two interstices A$_1$ and A$_2$, while in panel (\textit{c}), the two interstices have merged again. 

\begin{figure}
	\includegraphics{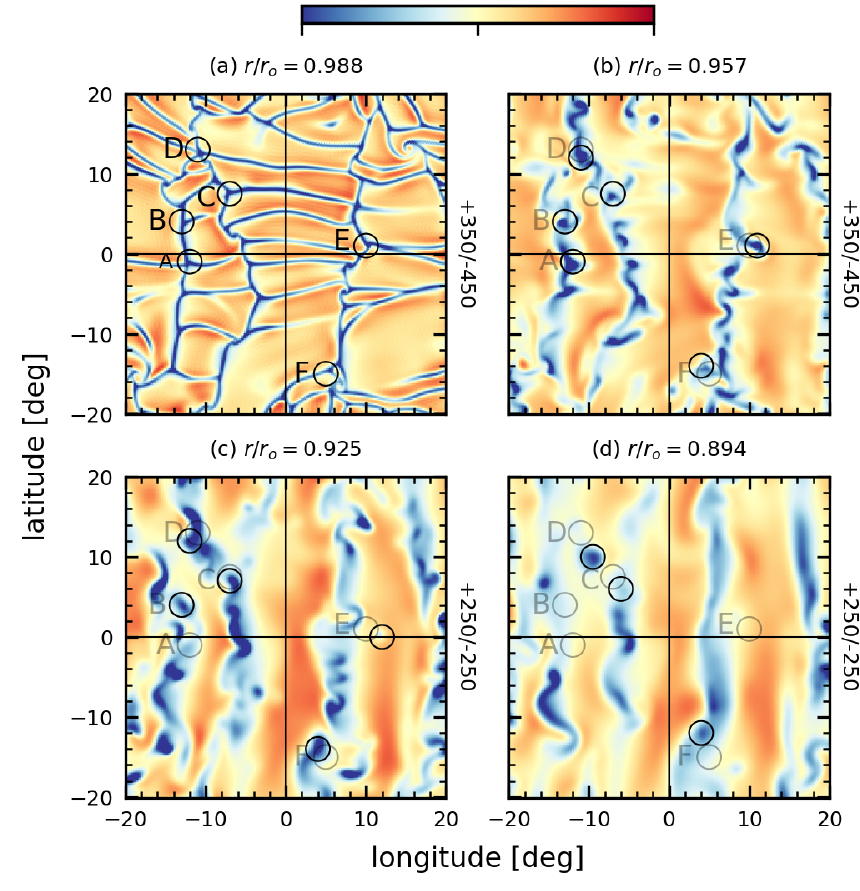}
	\caption{Magnified view of the $40^{\circ}\times40^{\circ}$ equatorial patch of radial velocity denoted by the dashed box in Figure \ref{fig:sslice}($c$). (\textit{a})--(\textit{d}) show the same patch at a common time for successively deeper layers, evenly spaced between $\sim$5\% depth and $\sim$50\% depth. The saturation values (in m/s) for the colorbar are shown to the right of each panel. Several downflow plumes have been traced in depth by small black circles, labeled in panel (\textit{a}) by capital letters A--F. Interstices A--C correspond to the same interstices identified in Figure \ref{fig:n5_3_patch_evolve}. The near-surface plume locations and labels are shown in gray in panels (\textit{b})--(\textit{d}). \label{fig:n5_vr_patch_dive}}
\end{figure}

At any given instant of time, the interstices shown in Figure \ref{fig:n5_3_patch_evolve} are the sources of \textit{downflow plumes}. The plumes can be seen by following the interstices down in depth. In Figure \ref{fig:n5_vr_patch_dive}, we magnify the $40^\circ\times40^\circ$ patch centered at the equator for case N5 and examine the connectivity of the downflows from the near-surface layers to mid-depth. As the patches get successively deeper, the downflow associated with the interstice intensifies in amplitude and becomes more localized. We refer to the entire structure (interstice to localized downflow at $\sim$mid-depth) as a \textit{downflow plume}. It is important to note that the plumes do \textit{not} coincide with the trajectory of a  fluid parcel launched downward from the interstice. Since the interstices super-rotate in time, each radial point on the plume corresponds to a fluid parcel that was launched when the parent interstice (top of the plume) was at a lower longitude.

The coherence of the North-South downflow lanes in Figure \ref{fig:n5_vr_patch_dive} increases with depth and the plumes slowly fade. No plume extends in depth more than $\sim$$0.1r_o$ (or about 2/5 the depth of the layer) from the near-surface layer shown in panel (\textit{a}).  This is consistent with the ephemeral nature of the interstices; over the radial extent of a plume, the fastest speeds are on average $\sim$400 m/s in the plume core, while the lifetime of an individual interstice is $\sim$2 days. Thus, the total depth-extent of the plumes should be $\sim (400\ \rm{m/s}) \times (2\ \rm{days})$ $\sim 0.1r_o$. 

\section{Torque Balance}\label{sec:torque}
The steady-state distribution of angular momentum (and by extension, the differential rotation) can be understood in terms of the angular momentum transport, or torque, due to various aspects of the flow. In equilibrium, the torque balance (e.g., \citealt{Elliot00}; \citealt{Brun02}; \citealt{Miesch11}) is expressed as
\begin{align}\label{eq:torque_balance}
\tau_{rs} + \tau_{mc} + \tau_v \equiv 0,
\end{align}
where
\begin{align}
\tau_{rs} &\equiv -\nabla\cdot [\overline{\rho} r\sin\theta \langle v^\prime_\phi \bm{v}^\prime_m\rangle]\nonumber\\
\tau_{mc}&\equiv -\langle\overline{\rho}\bm{v}_m\rangle \cdot\nabla\mathcal{L}\nonumber\\
\text{and}\ \ \ \ \ \tau_{v} &\equiv \nabla\cdot[\overline{\rho}\nu r^2\sin^2\theta\nabla\Omega]. \label{eq:torque_def}
\end{align}
Here, $\bm{v}_m \equiv v_r\hat{\bm{e}}_r + v_\theta\hat{\bm{e}}_\theta$ is the meridional part of the fluid velocity and $\mathcal{L}\equiv r\sin\theta(\Omega_0r\sin\theta + \langle v_\phi \rangle )=\Omega r^2\sin^2\theta$ is the fluid's specific angular momentum in the non-rotating lab frame. Angular brackets indicate a combined temporal and azimuthal average and the primes indicate deviations from the average. 

\begin{figure*}
	\includegraphics{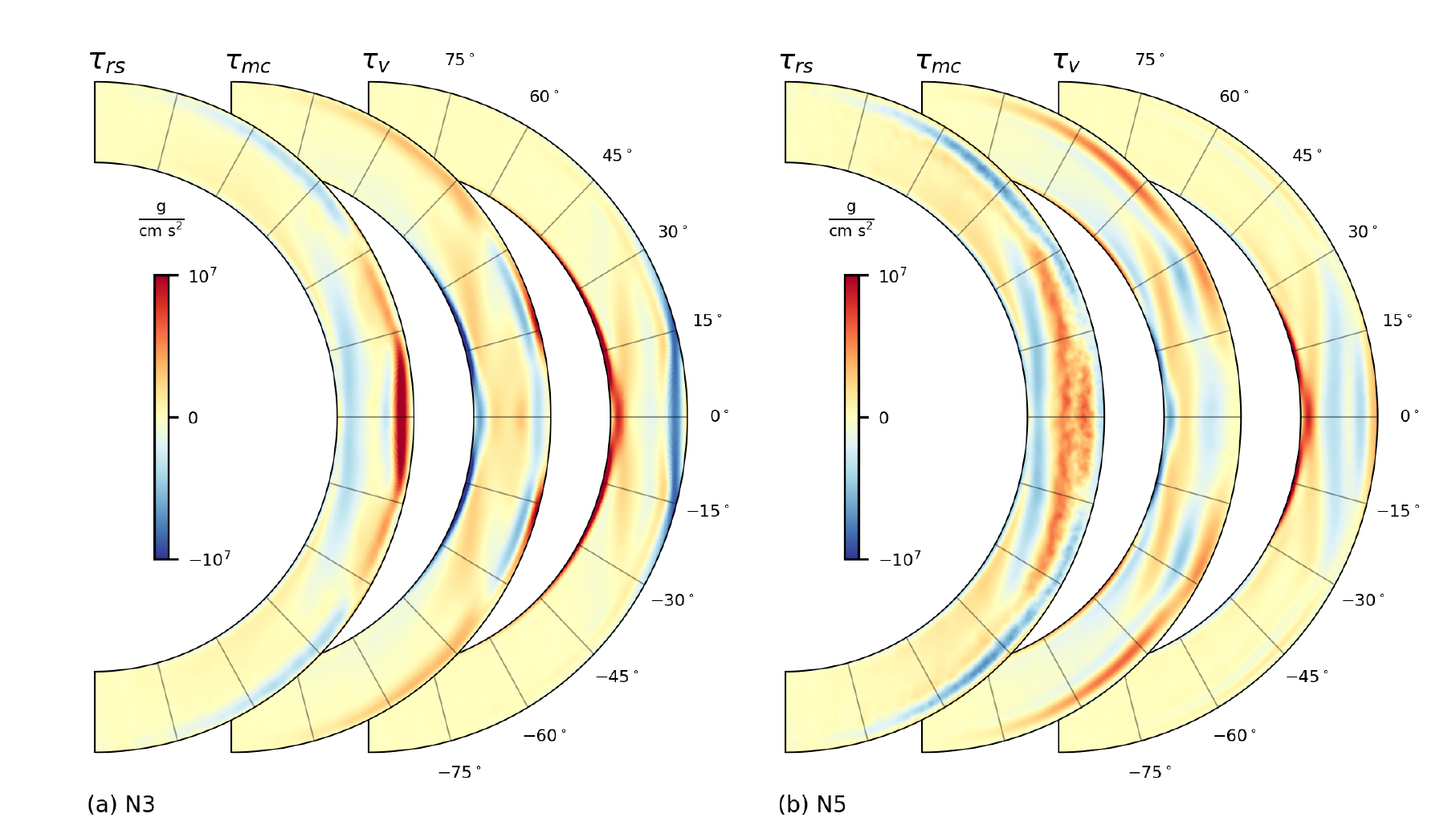}
	\caption{(\textit{a}) and (\textit{b}): Temporally and azimuthally averaged torque balance in the meridional plane for cases N3 and N5, respectively. In each panel, overlapping plots show the torque densities due to Reynolds stress, meridional circulation and viscosity from left to right. Positive torque, which tends to increase angular momentum in the axial direction $\hat{\bm{e}}_z$  (or tries to accelerate the fluid in the azimuthal direction $\hat{\bm{e}}_\phi$) is shown in red and negative torque in blue.  \label{fig:torques}}
\end{figure*}

Figure \ref{fig:torques} shows the balance of torques for cases N3 and N5. At each point in the meridional plane, the magnitude of the sum of the torques is a factor of $\sim$100 smaller than the magnitude of any of the three torques individually, indicating a mature state of equilibrium. For each case, the torque is roughly constant on cylinders near the equator, while it is roughly constant on spheres at higher latitudes. The main feature that sets case N5 apart from case N3 is the strong band of negative Reynolds stress torque near case N5's outer surface, extending across all latitudes; in case N3, there is only negative Reynolds stress torque at high latitudes, and it is significantly weaker than in case N5. 

A decomposition of $\langle v_\phi^\prime\bm{v}^\prime_m\rangle$  into its two component-correlations $\langle v_r^\prime v_\phi^\prime\rangle$ and $\langle v_\theta^\prime v_\phi^\prime\rangle$ reveals that the Reynolds stress torque is almost entirely dominated by the \textit{radial} turbulent angular momentum flux
\begin{align}
\mathcal{F}_r &\equiv \overline{\rho}r\sin\theta \langle v_r^\prime v_\phi^\prime\rangle
\label{eq:F_rs_r}
\end{align} 
for both cases N3 and N5.  We now argue that this radial flux is produced by the two types of flow structures mentioned previously, Busse columns and downflow plumes.

\begin{figure*}
	\includegraphics{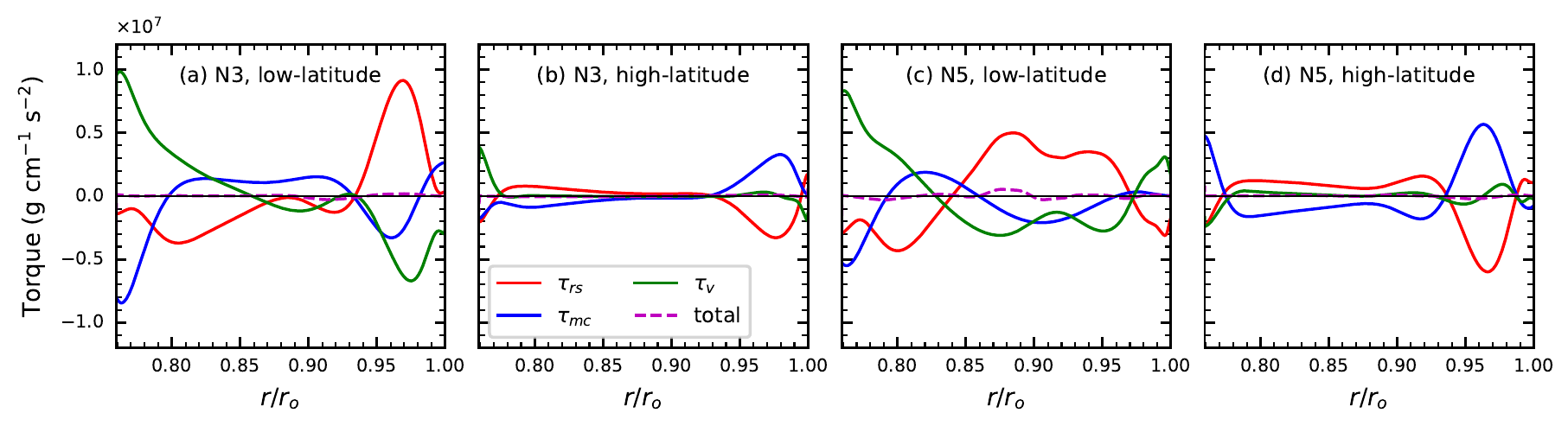}
	\caption{($a$)---($d$):  Torque balance averaged over low latitudes (between $\pm15^\circ$) and high latitudes (between $\pm45^\circ$ and $\pm60^\circ$) for cases N3 and N5. Each curve shows a torque averaged over its particular latitude regime and plotted as a function of radius. \label{fig:torques_rslice}}
\end{figure*}

The cross-sectional tilt of the Busse columns (Figure \ref{fig:eqslice}) tends to give upflows ($v_r^\prime>0$) a positive $v_\phi^\prime$ and downflows ($v_r^\prime<0$) a  negative $v_\phi^\prime$. Within a Busse column, both upflows and downflows thus yield a positive (radially outward) angular momentum flux. The net result is that angular momentum is taken away from the bottom layers (negative torque) and deposited in the upper layers (positive torque). Figure \ref{fig:torques_rslice}(\textit{a}) shows the low-latitude torque balance for case N3. The Reynolds stress torque is positive in the upper quarter of the domain and negative in the lower three-quarters, indicating that Busse columns dominate the Reynolds stress at all depths. In case N5, however, the Reynolds stress torque is negative near the outer surface, indicating the influence of an additional transport mechanism.

The downflow plumes do not follow the Busse column tilt. There are only a few other physical mechanisms to create correlations in $v_r^\prime$ and $v_\phi^\prime$, and the most obvious of these is deflection by the Coriolis force. For downflow plumes ($v_r^\prime<0$), the deflection would be prograde ($v_\phi^\prime>0$), corresponding to a negative (radially inward) transport of angular momentum. This picture is consistent with the Reynolds stress torque at high latitudes for both N3 and N5 (Figure \ref{fig:torques_rslice}((\textit{b}) and (\textit{d}))): the Reynolds stress torque is negative in the upper layers and positive in the deeper layers, corresponding to the inward transport of angular momentum from the outer surface to the bottom. 

At low latitudes, the presence of downflow plumes is most prominent in case N5 near the outer surface, where the Reynolds stress torque is negative. Correspondingly, there is a dip in rotation rate at low latitudes for case N5, but not for case N3. This lends substantial support to the argument of \citet{Foukal75}, at least at low latitudes: downflow plumes are Coriolis-deflected (or equivalently, they conserve their angular momentum), transporting angular momentum radially inward and creating near-surface shear. 

The torque from the meridional circulation is determined by the alternating pattern of clockwise/anticlockwise cells in Figure \ref{fig:massflux}, coupled with the fact that angular momentum in our models is roughly constant on cylinders ($\lambda$ = constant); from Equation \eqref{eq:torque_def}, this means that the torque has a sign opposite to that of $v_\lambda$. Near the equatorial boundaries of the circulation cells, the meridional flow is dominated by the radial component. Consequently, $\tau_{mc}$ changes sign in Figure \ref{fig:torques_rslice}((\textit{a}) and (\textit{c})) at the radial locations of the North--South cell boundaries.

At high latitudes, a decomposition of the meridional circulation into its radial and latitudinal components reveals that only the latitudinal term $-\langle\overline{\rho}v_\theta\rangle(\partial\mathcal{L}/r\partial \theta)$ contributes significantly to the torque. Geometrically, this is due to the shape of the near-surface anticlockwise cell of meridional circulation in both N3 and N5, which is most intense in a near-surface region highly elongated in the latitudinal direction. The resultant poleward flow dredges up high-angular-momentum fluid from the equator and brings it to high latitudes, thereby creating a positive near-surface torque.

The viscous flux of angular momentum ($-\overline{\rho}\nu r^2\sin^2\theta\nabla\Omega$) is proportional to the negative gradient of the rotation rate. This is similar to a Fickian diffusivity, meaning that viscosity simply tends to bring the rotation rate to a constant value. In this sense, the viscous torque is ``passive," responding only to counteract the shear produced by the Reynolds stress and meridional circulation. Thus, in Figure \ref{fig:torques_rslice}, the viscous torque is simply negative or positive according to the sign of the combined Reynolds stress and meridional circulation torque. 

\section{Angular Momentum Transport from Upflows and Downflows}
\label{sec:amom_flux}
\begin{figure*}
	\includegraphics{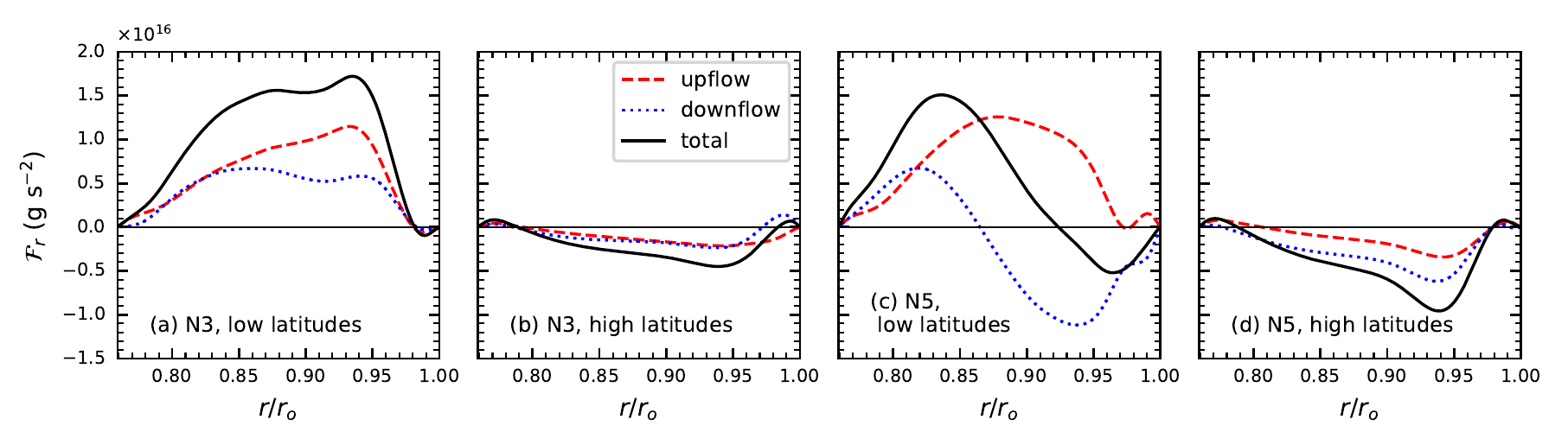}
	\caption{(\textit{a})--(\textit{d}): The radial angular momentum flux from the Reynolds stress ($\mathcal{F}_r$) averaged over low latitudes (between $\pm15^\circ$) and high latitudes (between $\pm45^\circ$ and $\pm60^\circ$) for cases N3 and N5. The flux has been divided into a component from the upflows ($v_r>0$) and a component from the downflows ($v_r<0$). When added, these components equal the total flux.  \label{fig:amomflux}}
\end{figure*}
Figure \ref{fig:amomflux} shows the Reynolds stress angular momentum flux broken up into upflow- and downflow-components for cases N3 and N5 in the high- and low-latitude regimes. For case N3 at low latitudes, both components of the flux have the same sign---positive in most of the domain except for a narrow region of very weak negative flux near the outer boundary. The strong positive fluxes are consistent with the tilts of the Busse columns in the equatorial plane, which makes the correlation $\langle v_r^\prime v_\phi^\prime\rangle$ positive for both upflows and downflows. In the middle layers, the magnitude of the upflow-flux is about twice as large as that of the downflow-flux. This asymmetry may be attributed to the fact that although \textit{most} downflows are part of a Busse column, there are also downflows whose speeds are large enough that they are really plumes that lack a negative $v_\phi^\prime$. The overall positive correlation $\langle v_r^\prime v_\phi^\prime\rangle$ is thus weaker for the downflows.

For case N5, the low-latitude upflow-flux looks rather similar to the upflow-flux in case N3, except that it peaks in the middle of the layer, as opposed to peaking in the upper half.  The downflow-flux in the bottom half of the layer ($r/r_o\lesssim 0.87$) is positive, consistent with the organization of both upflows and downflows into Busse columns  in the deeper layers. The top half of the shell, by contrast, has negative downflow-flux, whose maximum magnitude is over twice as great as the maximum magnitude of the positive downflow-flux. Thus, at low latitudes, the inward transport of angular momentum comes \textit{only} from the downflow plumes; the upflows are dominated by Busse columns. 

When the low-latitude upflow and downflow-fluxes are added, there is a positive slope in the radial profile of the total angular momentum flux $\mathcal{F}_r$. The effect on the Reynolds stress \textit{torque}, which scales like the radial derivative of $r^2\mathcal{F}_r$, is to produce the narrow region of negative torque near the outer boundary as seen in Figure \ref{fig:torques_rslice}(\textit{c}). This negative torque is responsible for maintaining the low-latitude near-surface shear against viscosity.

\begin{figure}
	\includegraphics[width=0.5\textwidth]{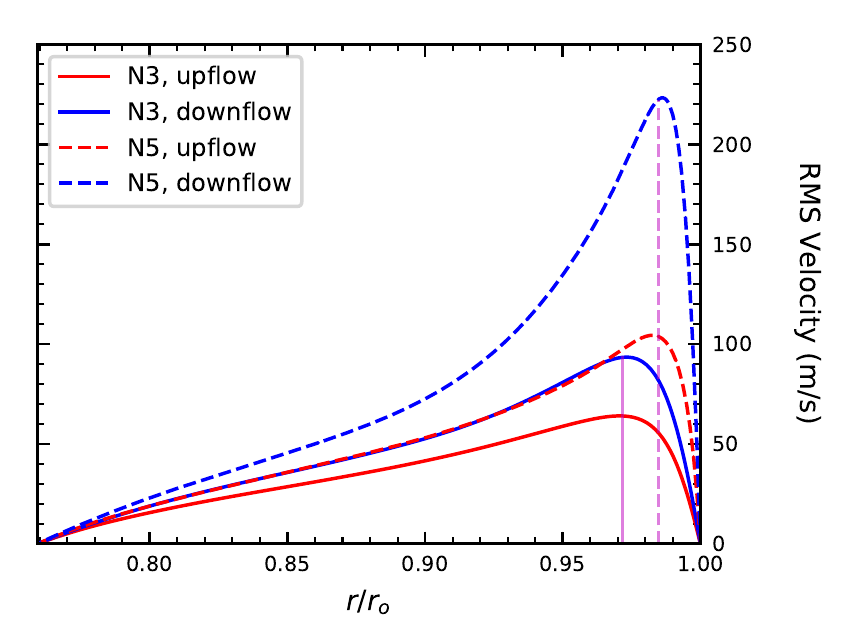}
	\caption{Spherically and temporally averaged radial convective velocity amplitudes for cases N3  and N5. The rms amplitudes are computed separately for upflows and downflows. The vertical magenta lines indicate the location of the base of each thermal boundary layer for cases N3 and N5 (compare to Figure \ref{fig:eflux_radial}).  \label{fig:vamp}}
\end{figure}

At high latitudes in both N3 and N5, the upflow- and downflow-fluxes have the same sign: negative in most of the fluid layer, except near the shell boundaries. Near the outer surface, there is significant asymmetry in the flux magnitudes between cases N3 and N5, and separately between the upflow- and downflow-flux of case N5. We argue that this arises mostly from the asymmetry in upflow- and downflow-speeds. Figure \ref{fig:vamp} shows the rms radial speeds of upflows and downflows in cases N3 and N5 as functions of radius. In case N5, the downflows are about twice as fast as the upflows near the outer surface. The Reynolds stress, which scales like $v_r^2$, is correspondingly larger for the downflows than the upflows. Similarly, since both the upflows and downflows are about twice as fast in case N5 than in case N3 near the outer surface, both the upflow- and downflow-fluxes are greater in case N5.

We note here that the Rossby number associated with the rms radial downflow speed at the base of the thermal boundary layer is $\sim$0.2 for case N3 and $\sim$1.2 for case N5. Here, we define the Rossby number through $\rm{Ro}_{\rm{c}}$ = $v_{rms}/2\Omega_0 H_\rho$, where $H_\rho$ is the local density scale height and $v_{rms}$ is the rms speed of the downflows. Hence, the \textit{downflows} in case N5 are, on average, rotationally \textit{unconstrained} according to Equation \eqref{eq:highRoc}. This agrees with the fact that the downflow-flux is negative in the outer layers of case N5; the downflow plumes conserve their angular momentum and thus transport it radially inward.

\begin{figure}
	\includegraphics{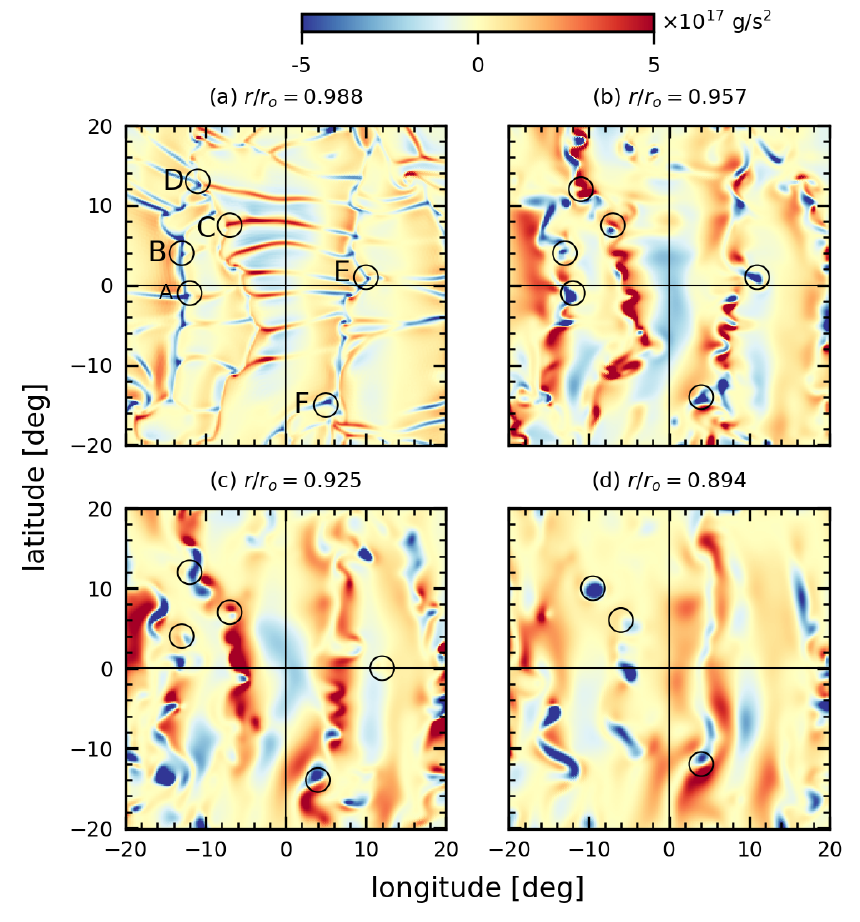}
	\caption{Magnified view of the same $40^{\circ}\times40^{\circ}$ equatorial patch of Figure \ref{fig:n5_vr_patch_dive}, this time showing the instantaneous angular momentum flux $\overline{\rho}r\sin\theta v_r^\prime v_\phi^\prime$ carried by each fluid parcel. Here, the colorbar is saturated to the same values for all four panels. The locations of the downflow plumes identified in Figure \ref{fig:n5_vr_patch_dive} are again traced in depth by small black circles, labeled in the first panel by capital letters A--F.  \label{fig:n5_vrvp_patch_dive}}
\end{figure}

We now verify that it is only the downflow \textit{plumes} that carry angular momentum inwards at low latitudes, and not the slower downflow lanes associated with Busse columns--which are rotationally \textit{constrained} according to Equation \eqref{eq:highRoc}.  Figure \ref{fig:n5_vrvp_patch_dive} shows a similar ``dive" through the fluid layer as in Figure \ref{fig:n5_vr_patch_dive}, but this time illustrates the instantaneous angular momentum flux in the patch.  We see that at $r/r_o=0.957$ (which is close to $r/r_o=0.942$---the extremum of the flux from the downflows in Figure \ref{fig:amomflux}(\textit{b})), the central region of nearly every plume is associated with a positive $v_\phi^\prime$ and thus a negative angular momentum flux. The notable exception is plume D, which has the deepest extent of any of the plumes. However, plume D has negative flux in the deepest layers. 

The Busse columns, by contrast, are mostly associated with positive angular momentum flux, consistent with their prograde tilt. It is also interesting to note that in between the axially aligned downflow lanes, the upflows are split into two regions with alternating sign of the flux---positive on the right and negative on the left. This is simply due to the manner in which upflows diverge and recirculate (in the center of the upwell) to accommodate the impenetrable outer boundary and maintain mass conservation. Note, however, that in the deeper layers, the positive flux in the upflows dominates the negative flux. This explains the weakness of the positive upflow-flux between $r/r_o\sim0.96$ and $r/r_o\sim1$ in Figure \ref{fig:amomflux}, panels (\textit{b}) and (\textit{d}). Near the top surface, the upflow rolls diverge symmetrically in both the positive and negative azimuthal directions and thus their net angular momentum transport cancels out almost completely. 

Finally, we note that near the outer surface, the East-West downflow lanes have largely the opposite sign of flux compared to the upflows in which they are embedded---negative on the right and positive on the left for the upflow column straddling the central meridian. This is due to the tendency of the fluid in the East-West lanes to flow sideways to the interstices, which have an extremely low pressure compared to their surroundings.

\section{Discussion}\label{sec:disc}
\begin{figure}
	\includegraphics{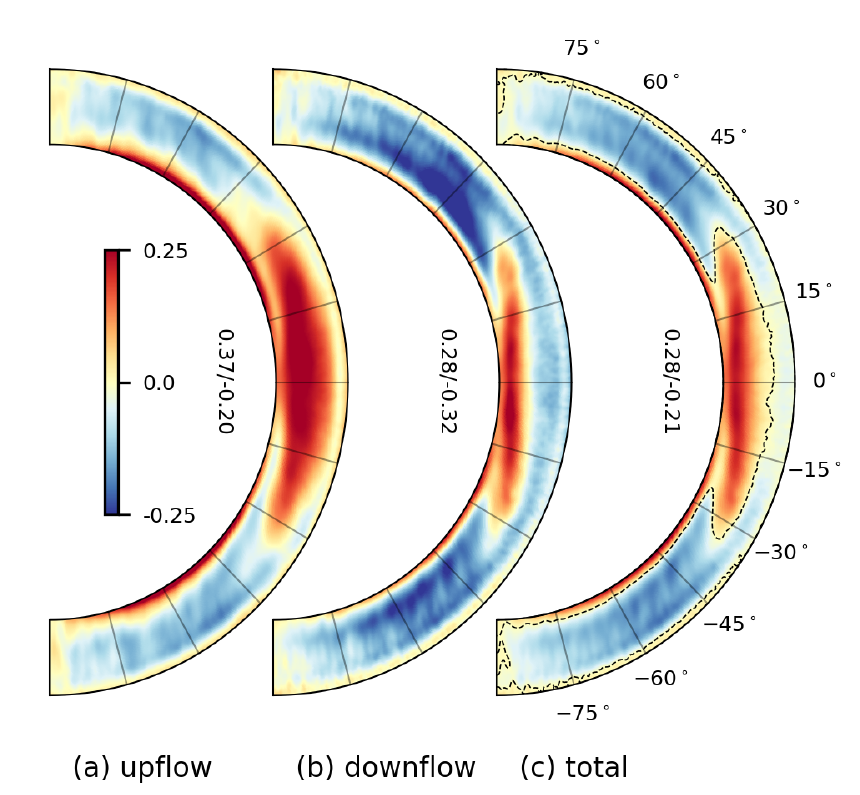}
	\caption{ (\textit{a}), (\textit{b}) and (\textit{c}) show the correlation coefficient of the convective velocities $v_r^\prime$ and $v_\phi^\prime$ for the upflows, downflows and total flow (respectively) in the meridional plane of case N5. The numbers beside each Figure indicate the extremum values for the  correlation coefficient. The dashed contour in panel (\textit{c}) denotes the surface in the meridional plane where the total correlation coefficient vanishes. \label{fig:n5_cor_vrvp}}
\end{figure}
In this work, we have explored the development of near-surface shear in 3D spherical-shell models of solar-like convection as a natural consequence of increasing the density contrast across the shell. We find that increased stratification does indeed foster more rapid flow structures with Reynolds stresses that enhance near-surface shear. However, this proves insufficient to create a solar-like near-surface shear layer. 

Our highest-contrast model (case N5) contains two types of flow structures that influence differential rotation: rotationally constrained Busse columns and rotationally unconstrained downflow plumes. The Busse columns transport angular momentum outward at low latitudes and thus maintain a fast equator and slow poles. The plumes transport angular momentum inward in the outer half of the layer at all latitudes. The influences of Busse columns and downflow plumes on the Reynolds stress is cleanly summarized in Figure \ref{fig:n5_cor_vrvp}, which shows the correlation coefficient $C=\langle v_r^\prime v_\phi^\prime\rangle/[\langle (v_r^\prime)^2\rangle\langle (v_\phi^\prime)^2\rangle]^{1/2}$ plotted in the meridional plane for the upflows and downflows in case N5. This correlation contains the same information as the angular momentum fluxes depicted in Figure \ref{fig:amomflux}, but the signal has been normalized and the geometric factor $r\sin\theta$ has been removed.

For the upflows, the correlation is everywhere positive at low latitudes, implying outward angular momentum transport---and thus dominance by Busse columns---at all depths. It is slightly negative (except near the inner boundary) at high latitudes, consistent with inward angular momentum transport through Coriolis-deflection. The correlation for the downflows is mostly negative everywhere, except in the inner half of the shell at low latitudes (where the Busse columns dominate) and near the inner boundary at high latitudes. The strong positive correlation near the inner boundary for both upflows and downflows is due to the impenetrability condition maintained by the pressure force, which we do not investigate in detail here.

Figure \ref{fig:n5_cor_vrvp} shows that the dip in angular velocity at low latitudes in case N5, while similar in amplitude and radial extent to the solar NSSL, arises for the wrong reasons. The Busse columns transport angular momentum outward, while the plumes transport angular momentum inward. The radial location where the two fluid structures meet (i.e., the spherical surface where the total correlation is zero in Figure \ref{fig:n5_cor_vrvp}(\textit{c})) corresponds roughly to the peak in rotation rate at low latitudes. The low-latitude NSSL in case N5 thus does not represent a layer that the downflow plumes slow down, but rather a layer that the Busse columns fail to speed up. In the real Sun, by contrast, the rotation curves have only a small positive slope with radius (see Figure \ref{fig:gongcut}), implying that Busse columns, if they are present in the Sun, do not transport angular momentum in the same way as they do in simulations, or that the plumes reach more deeply than in our models. 

The essential role played by Busse columns in maintaining numerical models' near-surface shear is even more apparent at high latitudes in case N5. Here, the Busse columns have little effect on angular momentum transport (the correlations in Figure \ref{fig:n5_cor_vrvp} are mostly negative for both upflows and downflows), while fast, small-scale downflow plumes conserve angular momentum in radial motion, transporting it inward. Although this Coriolis-deflection effect is, in fact, \textit{stronger} at high latitudes than at low latitudes,  there is basically no high-latitude near-surface shear in case N5. Thus, the mechanism described by \citet{Foukal75}---namely, that there is a rotationally unconstrained fluid layer near the outer surface of the convection zone---cannot, by itself, explain the NSSL, at least not using current numerical models. 

\citet{Miesch11} make a similar point in their investigation of how meridional circulation and differential rotation respond to a negative axial torque. In our case N5, there is a negative torque due to the Reynolds stress. At high latitudes, the torque balance is primarily between the Reynolds stress and meridional circulation, leading to the following equilibrium relationship between meridional circulation and differential rotation:
\begin{align}
\overline{\rho}\langle\bm{v}_m\rangle\cdot\nabla(\Omega r^2\sin^2{\theta})&=\tau_{rs}.\label{eq:inertial_balance}
\end{align}
The preceding equation may be satisfied in multiple ways, since both the rotation rate $\Omega$ \textit{and} the meridional circulation $\overline{\rho}\langle\bm{v}_m\rangle$ appear on the LHS. All local models that seek to explain the solar NSSL by inward angular momentum transport---for example, the Coriolis-deflection of downflow plumes in our case N5---prescribe a negative RHS, e.g., a negative $\tau_{rs}$. This does \textit{not}, however, constrain the rotation rate until another equation---namely, that of meridional force balance---is specified to determine the meridional circulation. Since the meridional circulation is fundamentally a global phenomenon, it is unlikely that any local model of angular momentum transport will explain the NSSL.

 An examination of the high-latitude meridional circulation profile in the context of Equation \eqref{eq:inertial_balance} reveals why near-surface shear is nearly absent at high latitudes in our case N5 and in the simulation of \citet{Hotta15}. In both simulations, there is a narrow band of poleward meridional circulation near the outer surface (see Figure \ref{fig:massflux} of this work and Figure 10 of \citealt{Hotta15}), which brings high-angular momentum fluid from equator to pole. If, on long timescales, this process pumps angular momentum to high latitudes more efficiently than the Reynolds stress can remove it, there will be no shear at high latitudes in equilibrium. 

\begin{figure}
	\includegraphics{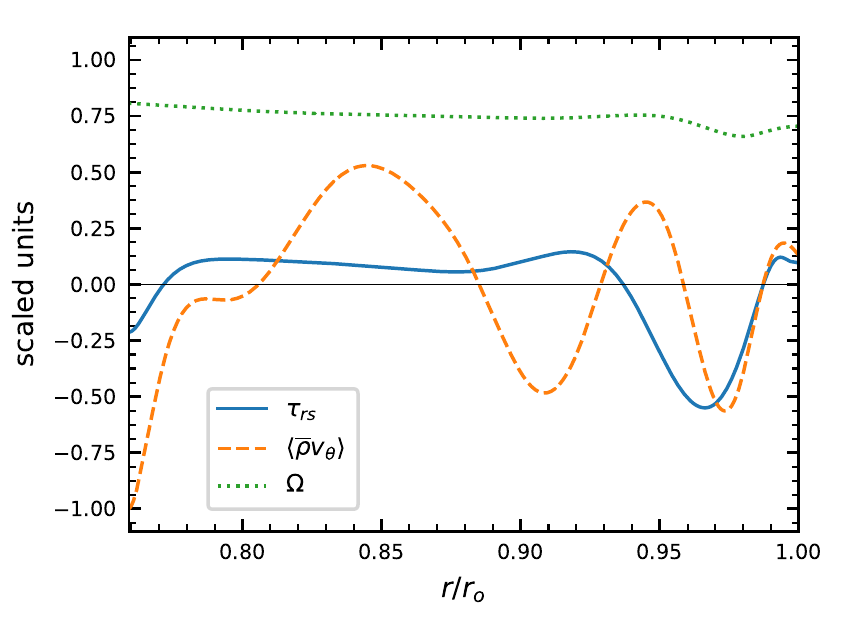}
	\caption{Radial profiles of $\tau_{rs}$, $\langle\overline{\rho}v_{\theta}\rangle$ and $\Omega$, averaged over high latitudes (between $\pm45^\circ$ and $\pm60^\circ$) for case N5. $\langle\overline{\rho}v_{\theta}\rangle$ has been scaled by its maximum absolute value to make its radial profile lie in the range $(-1, 1)$. $\Omega$ and $\tau_{rs}$ have been scaled to match the shape of their profiles in Figures \ref{fig:diffrot} and \ref{fig:torques_rslice}, respectively. $\Omega$ has been shifted to indicate that it is positive throughout the shell. \label{fig:highlat_profiles_n5}}
\end{figure}
Figure \ref{fig:highlat_profiles_n5} shows the radial profiles of Reynolds stress torque, latitudinal mass flux and rotation rate averaged over high latitudes for case N5. Clearly, case N5 satisfies Equation \eqref{eq:inertial_balance} by having the meridional circulation near the outer surface inherit the radial profile of negative Reynolds stress torque, while the rotation rate is mostly flat.

Because the molecular viscosity in the solar interior is so low, the balance in Equation \eqref{eq:inertial_balance} likely holds in the real Sun, possibly with the added complication of a Maxwell torque from the magnetic field. This work has shown explicitly that the only way to understand the solar NSSL in the context of local angular momentum transport---i.e., the RHS of Equation \eqref{eq:inertial_balance}---is through a detailed understanding of how the meridional circulation is established in the Sun. Once this is understood, the origins of the NSSL can be determined by Equation \eqref{eq:inertial_balance}. 

Although the detailed dynamical maintenance of the solar meridional circulation is unclear, helioseismic observations provide fairly rigorous constraints on the circulation profile itself, at least in the NSSL. In particular, the there is little near-surface variation of latitudinal flow $\langle \overline{\rho}v_\theta\rangle$ with radius in the Sun as compared to case N5 (e.g., \citealt{Giles97}; \citealt{Zhao04}; \citealt{Hathaway12}; \citealt{Chen17}; \citealt{Mandal18}). Given that both the meridional circulation $\langle \overline{\rho}\bm{v}_m\rangle$ and rotation rate $\Omega$ are observationally constrained in the NSSL, it would be worthwhile to measure the Reynolds stress torque in the NSSL as well, for example using high-resolution ring analysis (e.g., \citealt{Greer14}; \citealt{Greer15}). Once this is done, the relationship between meridional circulation and differential rotation in the NSSL will be elucidated. This will be an important guide for future simulations attempting to capture near-surface shear. 

In summary, we have determined that the solar NSSL is still an unsolved problem. The simple argument that a rotationally unconstrained layer near a convecting spherical shell's outer surface induces a NSSL does not hold up in convection simulations. This is due to the global character of the meridional circulation. Any local model of turbulent angular momentum transport thus provides an insufficient description of the dynamics of the NSSL. In future work, it would be useful to analyze the nature of the feedback between the meridional circulation and differential rotation achieved in spherical-shell models of convection, in particular its behavior in highly stratified regimes and in the presence of magnetic fields. Historically, an in-depth analysis of this feedback has rarely been done in simulations because the equilibrium meridional flow is achieved only over very long timescales (though see, for example, \citealt{Hotta15} and \citealt{Featherstone15}). This feedback is likely relevant in a broader context than simply NSSL dynamics. Since the meridional circulation is dynamically linked to the solar dynamo cycle both observationally and theoretically (e.g., \citealt{Wang89}; \citealt{Chou01}; \citealt{Ghizaru10}; \citealt{Charbonneau14}; \citealt{Komm15}), dissecting the nature of feedback between the solar NSSL and meridional circulation will provide important theoretical constraints on the processes by which the Sun forms its magnetic field.


\acknowledgments
We thank N. Featherstone for helping to formulate the idea of creating a rotationally unconstrained near-surface fluid layer in simulations by increasing density contrast, and also for his work as the primary developer of the \textit{Rayleigh} code. We thank K. Julien for helpful conversations on the subject of rotational constraint. We thank M. Miesch for illuminating discussion with regard to the high-latitude interaction between differential rotation and meridional circulation 

NASA grants NNX13AG18G and NNX16AC92G provided the primary support for this research. Additional support was provided through NASA grants NNX14AG05G, NNX14AC05G and NNX17AM01G.

Computational resources supporting this work were provided by the NASA High-End Computing (HEC) Program through the NASA Advanced Supercomputing (NAS) Division at Ames Research Center.

We thank the Computational Infrastructure for Geodynamics (\href{http://geodynamics.org}{http://geodynamics.org})---which is funded by the NSF under awards EAR-0949446 and EAR-1550901---for supporting the development of the \textit{Rayleigh} code. 

L. Matilsky was supported by a University of Colorado Boulder Chancellor Fellowship and a George Ellery Hale Graduate Fellowship.

\clearpage

\end{document}